\documentclass[referee,useAMS,esenatbib]{mn2e}
\usepackage{float}
\usepackage{graphicx}
\usepackage{booktabs}
\usepackage{subfigure}
\usepackage{amsmath}
\usepackage{amssymb}
\font\twelvei = cmmi10 scaled\magstep1
       \font\teni = cmmi10 
\font\mbf = cmmib10 scaled\magstep1
       \font\mbfs = cmmib10 \font\mbfss = cmmib10 scaled 833
\font\msybf = cmbsy10 scaled\magstep1
       \font\msybfs = cmbsy10 \font\msybfss = cmbsy10 scaled 833
\def\lsim{\raise0.3ex\hbox{$<$}\kern-0.75em{\lower0.65ex\hbox{$\sim$}}}
\def\gsim{\raise0.3ex\hbox{$>$}\kern-0.75em{\lower0.65ex\hbox{$\sim$}}}

\title[Effects of Photon Bending on the Comptonized Spectra of TCAF]
{Images and Spectral Properties of Two Component Advective Flows Around Black Holes: Effects of Photon Bending}
\author[Arka Chatterjee, Sandip K. Chakrabarti \& Himadri Ghosh]
{Arka Chatterjee\thanks{arka@csp.res.in}$^{1}$, Sandip K. Chakrabarti\thanks{chakraba@bose.res.in}$^{2,1}$ 
and Himadri Ghosh\thanks{himadri.ghosh@heritageit.edu}$^{3}$\\
$^{1}$Indian Centre for Space Physics, Chalantika 43, Garia Station Rd., 
	     Kolkata, 700084, India\\ 
$^{2}$S. N. Bose National Centre for Basic Sciences, Salt Lake,
              Kolkata, 700098, India\\
$^{3}$Heritage Institute of Technology, Kolkata, 700107, India}
\begin{document}

\date{}

\maketitle

\label{firstpage}

\begin{abstract}

Two component advective flow (TCAF) successfully explains spectral and timing 
properties of black hole candidates. We study the nature of photon trajectories 
in the vicinity of a Schwarzschild black hole and incorporate this in predicting 
images of TCAF with a black hole at the Centre. We also compute the emitted spectra. 
We employ a Monte-Carlo simulation technique to achieve our goal. For accurate 
prediction of the image and the spectra, null trajectories are generated without 
constraining the motion to any specific plane. Red shift, bolometric flux and corresponding 
temperature have been calculated with appropriate relativistic consideration. The 
centrifugal barrier dominated boundary layer or CENBOL near the inner region of the disk
which acts as the Compton cloud is appropriately modelled as a thick accretion 
disk in Schwarzschild geometry for the purpose of imaging and computing spectra. 
The variations of spectra and image with physical parameters such as 
the accretion rate ($\dot{m}_d$) and inclination angle are presented. We show that the
gravitational bending effects of photons do change the spectral shape to some extent.

\end{abstract}


%%%%%%%%%%%%%%%%%%%%%%%%%%%%%%%%%%%%%%%%%%%%%%%%%%%%%%%%%%%%%%%%%%%%%%%%

\section{Introduction}

Identification of any black hole candidate (hereafter BHC) involves in observing 
radiations from matter accreted on it. Radiations give out information on spectral 
and temporal properties in X-rays, UV or optical band. In near future, with the 
advent of high-resolution imaging devices, imaging of such BHCs with 
matter all around will be possible. It is therefore essential to follow paths of 
photons emitted from hydrodynamically correct matter distribution with successful 
spectral features very accurately till they reach observer's image plane. This 
would enable us to obtain the spectra and images from different inclination 
angles. In this paper, we achieve this by taking a two component advective 
flow (TCAF) solution based on rigorous theory of transonic astrophysical flows 
(Chakrabarti, 1990) and computing emitted radiation by a Monte-Carlo simulation. 

Bending of photons in a gravitational field is a natural consequence of general 
theory of relativity. In our context, photons emitted from the TCAF will be deviated 
by the strong gravitation field created by the black hole. Deviations of photons 
depend on the space-time metric and its impact parameter. Due to focusing effects, 
softer photons from the Keplerian component may be forced to intercept hot electrons 
in the post-shock region of the low-angular momentum component located close to the 
horizon modifying the spectrum. Theoretical work on imaging of emitting matter 
surrounding a black hole started in 1970s. The first step towards this emitter-to-observer 
ray tracing was attempted by Cunningham \& Bardeen (1973). Isoradial curves 
were drawn for a rotating black hole at some specified inclination angle. Pineault 
\& Roeder (1977) repeated this work with the variation of spin parameter. A 
complete image of a black hole surrounded by rotating matter was first drawn by 
Luminet in 1979. Idealized disk particles with differential rotational 
velocity were kept on the equatorial plane to construct a Keplerian disk. Mass of the 
disk was considered to be negligible as compared to the mass of the black hole. 
Emission of photon was assumed to be isotropic in nature. Bolometric flux of the 
Keplerian disk and red-shift of photons were included in the image. The basic process 
of generating an image was considered to be a boundary value problem, connecting 
a disk and an observer. In that way, the transfer function using elliptic 
integral is a time saver. Fukue \& Yokoyama (1988) produced a color photograph 
of a multi-wavelength study of the disk. They also studied flux variation in an 
eclipsing binary. Karas et al. (1992) developed a catalogue of photon trajectories 
with variable inclination angle and spin. Viergutz (1993) first attempted emitter-to-observer 
problem in the Kerr metric by generalizing the transfer function with rotating geometry. 
Until this work, images were drawn using semi-analytical methods, by computing 
a relation between the impact parameter ($b$), polar angle ($\alpha$) on observer 
plane and observer's polar angle ($\theta_{\circ}$). Marck (1996) integrated each 
trajectory of photons to generate an image for the first time. We have followed this 
approach in this paper. However, instead of using first order differential equations, 
we traced the ray using three dimensional second order equations to draw the complete 
photon trajectory. This gives us the freedom to use analytically established as well 
as time dependent three dimensional disks emitting photons with all known phase-space 
coordinates (initial position and momentum of photons) as a starting configuration.   

In twenty-first century, the use of X-ray polarimetry flourished rapidly. With that, 
Bromley et al. (2001) added polarization while drawing the disk image. Armitage and 
Reynolds (2003) performed magneto-hydrodynamics (MHD) on Keplerian disk and provided 
temporal variations of a Keplerian disk. M\"{u}ller \& Frauendiener (2011) developed 
Figures using the same transfer functions. `GYOTO' (Vincent et. al. 2011) is an 
open software available for imaging and trajectory calculations. More recently,
`YNOGK' (Yang \& Wang 2013), another publicly available code for Kerr space time was 
published. However, the present literature is mostly dominated by the images of a most ideal 
disk configuration, namely, a Keplerian disk, which unfortunately does not describe 
all the spectral states accurately. More recently, Younsi et al. (2012) published 
images containing relativistic thick disks for different inclinations and spin 
parameter with variation of gas pressure to radiation pressure ratios. However, 
so far, study of disk configurations, such as TCAF (Chakrabarti 1990; Chakrabarti 
\& Titarchuk, 1995; hereafter CT95), in different spectral states have been missing.

Since a black hole is very compact, the resolving power of a telescope must be very 
high to capture images close to the horizon. So, imaging by the so-called Event Horizon 
Telescope concentrates on nearby relatively massive black holes such as those present at the centre 
of our galaxy, namely, Sgr A* (Johnson et al. 2015). However, so far, no major attempt 
has been made to study the difference in images when the BHC is in the soft state and 
in the hard state, i.e., changes in the configuration of the disk itself. 
Part of the reason is that the model builders of spectral properties 
were keeping a Keplerian disk as the basis of all such spectral states and only the 
randomly placed Compton clouds (clouds of hot electrons) would change its properties. 
The hot cloud is required for inverse Compton scattering (Sunyaev \& Titarchuk, 1985). 
A concrete understanding of where the Compton cloud is really located and how its geometry 
vis-a\'-vis the Keplerian disk should change with spectral states came about only 
after successful application of the transonic flow solution in the context of black 
hole accretion by CT95. This so-called two component advective flow (TCAF) solution
is regularly applied to explain spectral evolutions of numerous outbursting BHCs
(see, Debnath et al. 2014; Mondal et al. 2014; Jana et al. 2015; Molla et al. 2016; 
Debjit et al. 2016) and actual physical parameter such as the accretion rates in the 
two components, the compression ratio of the centrifugally supported shock structure 
and the size of the Compton cloud itself were derived. According to CT95, state 
transitions are due to changes in accretion rates, only the time scale of such changes
varies with the mass of the black hole. Higher accretion rates in the 
Keplerian component relative to the advective component cools the hot electrons in 
the Compton cloud shrinking it to a small size resulting in a softer state. In the 
opposite case of higher advective flow rate, Compton cloud (post-shock region) remains 
hotter and a hard state is produced with a truncated Keplerian disk with its inner 
edge coinciding with the Compton cloud boundary or the shock location. So, geometry 
of the flow drastically changes with the spectral state. This is not common in massive black holes
but there are certainly active galaxies with both classes of spectra. In stellar mass black holes,
this change of geometry is common and quicker. For instance, in GRS 1915+105, there 
are more than dozens of variability classes and all these are physically distinguished 
by the system geometry governed by the size of the Compton cloud and the truncated 
Keplerian component (Pal \& Chakrabarti, 2011, 2013).  

One of the important results by an earlier numerical simulations by Molteni et 
al. (1994) is that the post-shock region (otherwise called the CENtrifugal pressure 
supported BOundary Layer or CENBOL, used as the Compton cloud in TCAF of CT95) of 
a transonic flow has most of the characteristics of a thick accretion torus (Abramowicz 
et al., 1978 and Kozlowski et al., 1978; Paczy\'nski \& Wiita, 1980; Begelman. 
Blandford \& Rees, 1982; Chakrabarti, 1985, hereafter C85), except that CENBOL 
also has a radial motion and thus more accurate than the thick disk models of 
early eighties mentioned above. In an attempt to have a stationary image we require 
a theoretical solution for the CENBOL. We therefore use the fully general relativistic 
thick torus solution of Chakrabarti (1985) as the CENBOL along with a truncated Keplerian 
disk to have a theoretical model of a two component advective flow. To have different 
spectral states we change the accretion rates of these two components. To generate 
the spectrum from it, we employ a Monte-Carlo simulation code where soft photons 
from the truncated Keplerian component are allowed to (inverse) Compton scatter 
from the relativistic Maxwellian electrons inside the CENBOL. Monte-Carlo process 
(Pozdnyakov et al., 1983) provides one of the best techniques to generate the complete 
spectrum. The power-law component in the observed spectrum of black hole candidates 
is the result of repeated inverse Comptonization (Sunyaev \& Titarchuk, 1985) of the 
soft photons. Thus, combining thermal multi-colour black body photons from the Keplerian 
disk, with the power-law component from thermal Comptonization, one obtains a complete 
spectrum. The Comptonized photons then follow curved trajectories as they leak out 
from the CENBOL after possible repeated scatterings and reach the observer. Higher 
energy X-rays are expected to come from deep inside of the CENBOL after multiple 
scatterings. Thus, one can obtain a different image with photons of different energy range. 
This makes the problem very interesting as far as observations are concerned. From 
the same simulation, by filtering photons reaching out to a series of observers sitting at 
different inclination angles, we also obtain totally different images and spectra 
of the same source with the same intrinsic disk properties. The time lag properties 
of this configuration would be important also. This would be reported elsewhere.

The plan of our paper is the following. In the next Section, we discuss geometry 
of the CENBOL and the Keplerian disk in our simulation. In \S 3, we describe the 
equations governing the null trajectories and image generation process that were 
used in this work. In \S 4, we present results of Monte-Carlo simulations for 
different disk parameters and in \S 5, we draw conclusions.

\section{Nature of the Compton cloud and the soft photon source}

In TCAF scenario, the higher viscosity component forms a Keplerian disk while the 
lower viscosity and low angular momentum component (which surrounds the Keplerian flow) 
forms a centrifugal barrier close to a black hole. In a large region of the parameter 
space, the centrifugal barrier induces formation of a standing or an oscillatory shock. 
The post-shock region is the Compton cloud. Being subsonic and basically rotating 
with more or less constant angular momentum, CENBOL satisfies some of the conditions 
of a radiation pressure supported thick accretion disk or a hot ion pressure supported 
torus (Abramowicz et al., 1978; Kozlowski et al., 1978; Paczy\'nski \& Wiita, 1980; 
Begelman, Blandford \& Rees, 1982; Chakrabarti, 1985, hereafter C85) and therefore 
for all practical purposes, one can use C85 prescription of `natural angular momentum' 
in general to describe electron number density and temperature distribution inside a 
hot CENBOL. CENBOL remains a hot ion pressure supported torus when the Keplerian 
component has a little accretion rate and the number of soft photon is not enough to cool 
down the CENBOL. We follow the procedure given in C85 to obtain the temperature and density 
distribution inside the CENBOL and treat it as the Compton cloud. In TCAF, the outer 
edge of the CENBOL is also the inner edge of the Keplerian component. Soft photons 
from the Keplerian disk is intercepted by the CENBOL and are re-emitted after Comptonization. 
In very soft states, when the Keplerian component rate is high, the CENBOL collapses 
and the entire disk can be treated as the standard Shakura-Sunyaev (1973, hereafter SS73) 
type disk which will pass through the inner sonic point due to its nature of transonicity 
(Chakrabarti, 1990). With even higher accretion rates, the inner edge is puffed up 
to produce a radiation pressure supported torus, but we will not explore that regime 
here. In Fig. 1, we present the geometry of our simulation. In order to obtain a 
stationary solution, we replace the high viscosity component of TCAF by the standard  
Keplerian disk as suggested by Page \& Thorne (1974) (relativistic counterpart of SS73) 
and the advecting CENBOL by a purely rotating thick ion torus (C85). Typical schematic 
photon trajectories are shown.

\begin{figure}
\centering
\vbox{
\includegraphics[width=14.0cm,height=5.0cm]{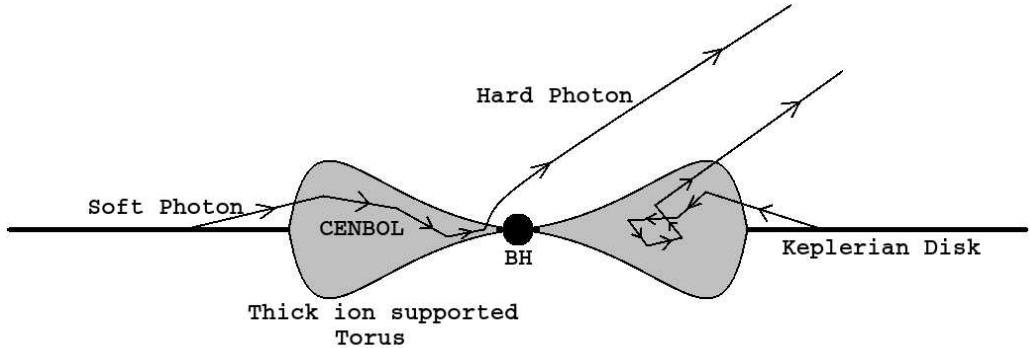}}
\caption{Our simulation geometry of two component advective flow solution consisting 
of one high viscosity truncated Keplerian disk component (Page \& Thorne 1974)
and one low angular momentum Polish donut type thick ion supported torus component (C85). 
Typical photon trajectories of soft and hard photons are shown. The advective component 
before the shock outside CENBOL is tenuous and is ignored.}
\end{figure}

\subsection{CENBOL as a relativistic torus}   

To describe CENBOL, we note that the relativistic Euler's equation for the perfect 
fluid under the influence of gravity, centrifugal force and pressure gradient force 
can be written as (C85),
\begin{equation}
\frac{\nabla p}{p+\epsilon} = -\ln(u_{t}) + \frac{\Omega \nabla l}{(1-\Omega l)}
\end{equation}
where, $p$ is the pressure and $\epsilon$ is the total energy density. Also, $l=-u_{\phi}/u_{t}$ 
is specific angular momentum and $\Omega =u^{\phi}/u^{t}$ is the relativistic angular 
momentum. Velocity components $u^{r}$ and $u^{\theta}$ are negligible for the present 
scenario. We consider barotropic process $p=p(\epsilon)$ such that constant pressure 
surfaces and equipotential surfaces coincides. Then, Euler's equation becomes,
\begin{equation}
W-W_{in} = \int_0^p\frac{dp}{p+\epsilon}=\int_{{u_{t}}_{in}}^{u_{t}} \ln(u_{t}) - \int_{l_{in}}^{l}\frac{\Omega \nabla l}{(1-\Omega l)},
\end{equation}
where, for a given angular momentum distribution $\Omega=\Omega(l)$, we can easily 
integrate the third term of this equation. 

The ratio $l/\Omega=\lambda^2$ has dimension $[L^2]$. Choosing $l=c\lambda^n$ (C85), 
where $c$ and $n$ are constants, yields the von-Zeipel relationship as follows,
\begin{equation}
\Omega = c^{2/n}l^{1-2/n},
\end{equation}
and $\lambda=\frac{r\mathrm{sin}\theta}{(1-\frac{2}{r})^{1/2}}$ (C85). In our 
simulation, we choose $G=1$,~$M=1$ and $c=1$ such that $r_g=2$. 

Constant parameters $c$ and $n$ can be derived from the following conditions (C85):
\begin{equation}
\begin{aligned}
\displaystyle c\lambda^{n}_{in} = l_{k}(r=r_{in},\theta=\pi/2)~\mathrm{and} ~\\
\displaystyle c\lambda^{n}_{c} = l_{k}(r=r_{c},\theta=\pi/2),\\
\end{aligned}
\label{eq:xdef}
\end{equation}
where, the subscript $in$ and $c$ correspond to the quantities at the disk inner 
edge and disk center. We kept $r_{in}=4.0$ and $r_{c}=9.8$ to generate 
the innermost thick toroidal shell. Variation of the potential keeping $r_{in}$ 
and $r_{c}$ fixed gives a large number of shells which create the complete stationary 
CENBOL. Comptonization occurs inside this region. Since, electrons inside the 
disk generally have higher energy than the soft photons injected into it, the later 
are repeatedly inverse Comptonized before they leave the CENBOL and reach the observer 
in curved trajectories.

\subsection{Distribution of temperature and density inside the Compton cloud}

\begin{figure}
\centering
\vbox{
\includegraphics[width=.4\textwidth,angle= 0]{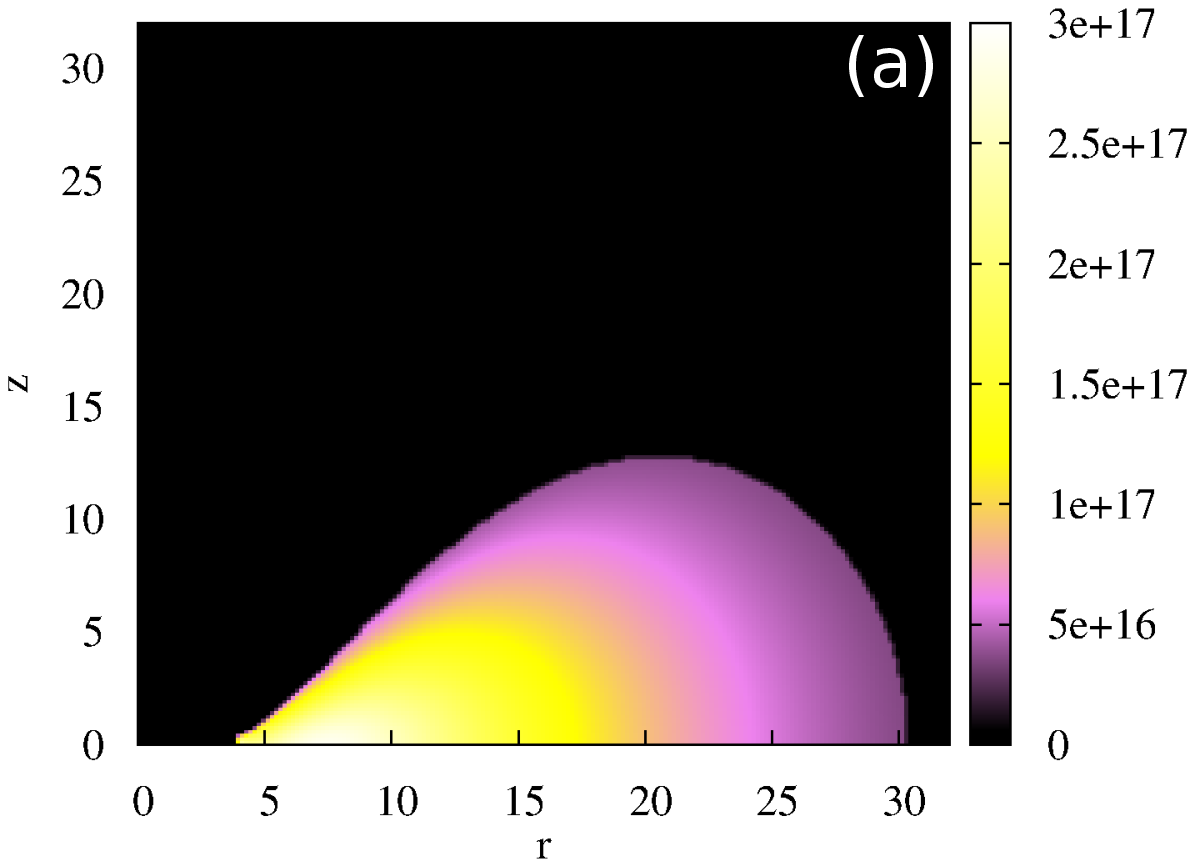}
\includegraphics[width=.4\textwidth,angle= 0]{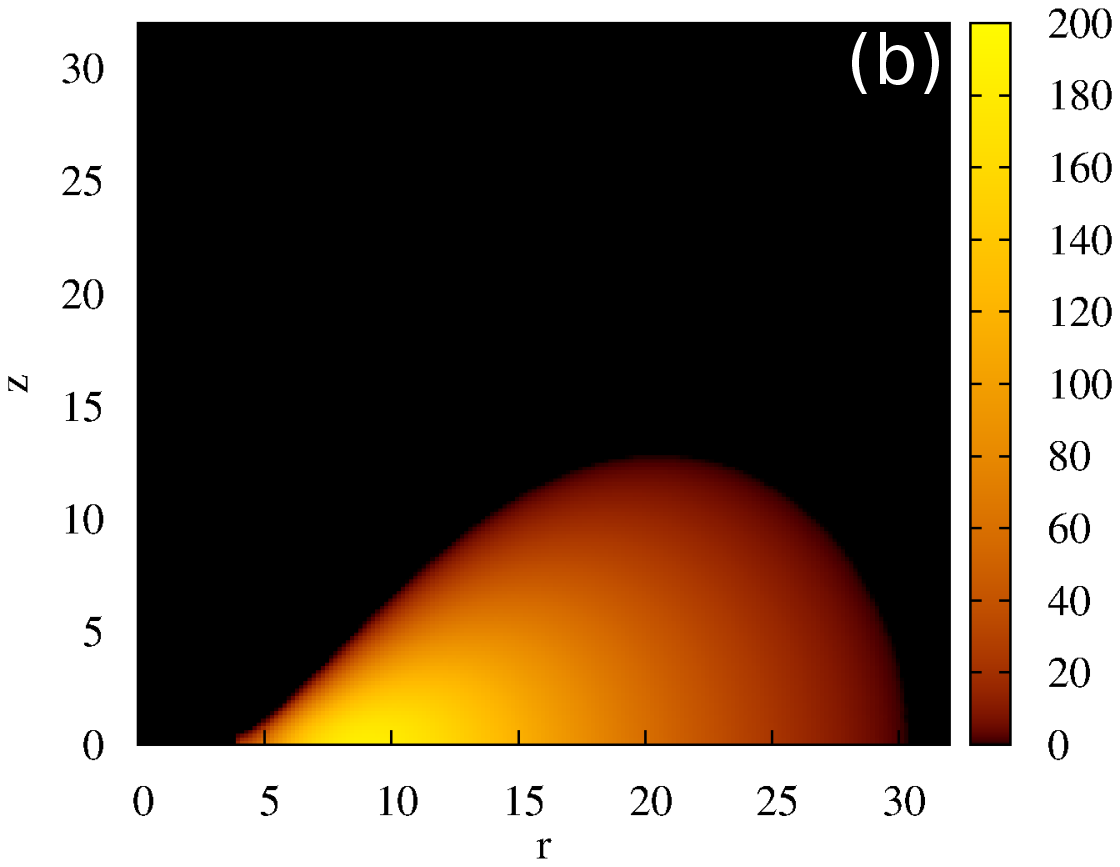}}
\caption{(a) Number density and (b) temperature profile of the CENBOL with inner edge at 
$4.0$, center at $9.8$ and outer edge at $30$. Temperature is given in keV
unit. Pressure, number density \& temperature of the outermost layer is kept at zero.}
\end{figure}

For Comptonization, we need the information of density, pressure and temperature inside
the CENBOL region. For this, we choose a barotropic equation of state where iso-temperature 
and iso-density surfaces coincide with equipotential surfaces. We calculate the disk 
temperature using a polytropic equation of state, $p=K\rho^{\gamma}$ where $p$, $\rho$, 
$K$ and $\gamma$ are pressure, matter density a constant related to entropy and polytropic 
index respectively. In a relativistic flow, one can choose $\gamma=4/3$. Since we 
are interested to study the effects of gravitational bending, we fix the CENBOL 
properties such that at $r=r_c$, the temperature is  $\sim 200$keV and the highest 
electron number density is $\sim 3.0\times10^{17}$. These can be achieved using a suitable 
entropy constant, $K$ as $T \propto \rho^{1/3}$. In Figure 2 we have shown the variation of 
electron number density (per $cm^3$) and temperature (in keV) inside the CENBOL.

\subsection{Soft Photon Source}  

Keplerian disk acts as the source of soft photons. We consider the disk outer edge to be 
only up to $50$ so as to save computing time. The inner edge is truncated just outside 
the CENBOL boundary. We assume this disk to be sitting on the equatorial plane. We use the 
simplified radiation profile as a function of radius as given by Page \& Thorne (1974). 
Bolometric flux of radiation from a Keplerian disk around a Schwarzschild black hole is written as: 
\begin{equation}
F_{k}^{disk}(r) = \frac{F_c(\dot{m}_d)}{(r-3)r^{5/2}}\bigg[\sqrt{r}-\sqrt{6}+
\frac{\sqrt{3}}{3}log\bigg(\frac{(\sqrt{r}+\sqrt{3})(\sqrt{6}-\sqrt{3})}
{(\sqrt{r}-\sqrt{3})(\sqrt{6}+\sqrt{3})}\bigg)\bigg],
\end{equation}
where, $F_c(\dot{m}_d)=\frac{3m\dot{m}_d}{8\pi r_{g}^3}$, with $\dot{m}_d$ being the disk
accretion rate in Eddington unit. From Stefan-Boltzmann law, one can easily get local 
temperature of the disk by,
\begin{equation}
T_{k}^{disk}(r)= \bigg(\frac{F_{k}^{disk}(r)}{\sigma}\bigg)^{1/4},
\end{equation}
where, $\sigma=\frac{2\pi^5k^4}{15h^3c^3}$ is the Stefan-Boltzmann constant.
Total number of photons emitted from Keplerian disk surface is given by
\begin{equation}
n_{\gamma}(r)=\bigg[16\pi \bigg(\frac{k_b}{hc}\bigg)^3 \times 1.202057\bigg](T_{k}^{disk}(r))^3
\end{equation}
The number of photons emitted from the radius $r$ to $r+\delta r$ is given by 
\begin{equation}
dN(r) = 4\pi r\delta rH(r)n_{\gamma}(r),
\end{equation}
where Keplerian disk height $H(r)$ is assumed to be $0.1$ (as $\frac{H(r)}{r} << 1$ for
Keplerian disk region). 
The Keplerian disk is divided in different annuli of width $D(r) = 0.1$. Each annulus 
having mean radius r is characterized by its average temperature $T_{k}^{disk}(r)$. 
The total number of photons emitted from the disk surface of each annulus can be
calculated using Equation (8). This total number comes out to be $\sim 10^{39}$ --$10^{40}$ per 
second for $\dot{m}_d=0.1$. One cannot inject this many photons in Monte Carlo 
simulation because of the limitation of computation time. So we replace this large 
number of photons by a lower number of bundles of photons, $N_{comp}(r) = 10^9$, and 
calculate a weightage factor $f_W=\frac{dN(r)}{N_{comp(r)}}$.

For different annulus, the number of photons in a bundle will change. From the 
standard disk model of Page and Throne (1974), we calculate $dN(r)$ and use that 
to calculate the change in energy via Comptonization. During inverse-Comptonization 
(or, Comptonization) by an electron in an elemental volume of $dV$, we consider that 
$f_W$ number of photons has suffered the same scattering with the electrons inside 
the same volume (Garain, Ghosh \& Chakrabarti 2012).

Monte-Carlo code randomly generates soft photons from the Keplerian disk.
Soft photon energy is calculated using Planck distribution law for a given
$T_{k}^{disk}(r)$. The number density of photons ($n_\gamma(E)$) having an
energy $E$ is given by,
\begin{equation}
n_\gamma(E) = \frac{1}{2 \zeta(3)} b^{3} E^{2}(e^{bE} -1 )^{-1}, 
\end{equation}
where, $b = 1/kT_{k}^{disk}(r)$ and $\zeta(3) = \sum^\infty_1{l}^{-3} = 1.202$,
the Riemann zeta function (e.g., Ghosh, Chakrabarti \& Laurent, 2009).

\subsection{Simulation Procedure}
To reduce the Computational time we consider the trajectories of photons to be 
straight lines {\it inside} the CENBOL. Difference of results with null geodesic 
path in between scatterings and straight line path was shown to be insignificant  
(Laurent \& Titarchuk, 1999). The injected soft photons are generated from the Keplerian 
disk at a random direction which required use of three random numbers. In the 
Monte Carlo simulation, the intensity of emitted radiation from the Keplerian 
disk $\propto \mathrm{cos}\theta$. That implies the Keplerian disk will
radiate most along the Z-direction and least along the equatorial 
plane. 

Another random number gives a target optical depth 
$\tau_c$ at which the scattering occurs. The photon is tracked within 
the electron cloud till the optical depth ($\tau$) reached $\tau_c$. 
The probability density of Compton scattering of a photon with an electron 
$\propto e^{-\tau}$, where $\tau$ is the optical depth of the cloud at that 
particular position. We calculate $\tau_c$ by randomizing this particular 
function using Monte Carlo technique. Finally, we get $\tau_c = -ln(\zeta)$, 
where $\zeta$ is a uniform random number between $0$ and $1$. The increase in 
optical depth ($d\tau$) as the photon travels a path of length $dl$ inside 
the electron cloud is given by: $d\tau = \rho_n \sigma dl$, where $\rho_n$ 
is the  local electron number density.

The total scattering cross section $\sigma$ is given by Klein-Nishina formula:
\begin{equation}
\sigma = \frac{2\pi r_{e}^{2}}{x}\left[ \left( 1 - \frac{4}{x} - \frac{8}{x^2} \right) 
ln\left( 1 + x \right) + \frac{1}{2} + \frac{8}{x} - \frac{1}{2\left( 1 + x \right)^2} \right],
\end{equation}
where, $x$ is given by,
\begin{equation}
x = \frac{2E}{m c^2} \gamma \left(1 - \mu \frac{v}{c} \right),
\end{equation}
$r_{e} = e^2/mc^2$ is the classical electron radius and $m$ is the mass of the electron.

We assumed here that a photon of energy $E$ and momentum $\frac{E}{c}\bf{\widehat{\Omega}}$
is scattered by an electron of energy $\gamma mc^{2}$ and momentum 
$\overrightarrow{\bf{p}} = \gamma m \overrightarrow{\bf{v}}$, 
with $\gamma = \left( 1 - \frac{v^2}{c^2}\right)^{-1/2}$ 
and $\mu = \bf{\widehat{\Omega}}. \widehat{\bf{v}}$. 
At this point, a scattering is allowed to take place. The photon selects an electron 
and the energy exchange is computed using 
the Compton or inverse Compton scattering formula. The electrons
are assumed to obey relativistic Maxwell distribution inside the Compton cloud.
The number $dN(p)$ of Maxwellian electrons having momentum between
$\vec{p}$ to $\vec{p} + d\vec{p}$ is expressed by,
\begin{equation}
dN(\vec{p}) \propto exp[-(p^2c^2 + m^2c^4)^{1/2}/kT_e]d\vec{p}.
\end{equation}

While travelling from one scattering centre to another, the gravitational redshift changes the
frequency of the photon. The process continues until the photon leaves CENBOL region or is 
sucked in by the black hole. The process is similar to what was used in Ghosh, 
Chakrabarti \& Laurent (2009), Ghosh et al. (2010, 2011), Garain et al. (2012), 
(2014); which was adopted originally from Pozdnyakov, Sobol and Sunyaev (1983).
We choose the mass of the black hole ($M=10 M_{\odot}$) in our simulation. In
Fig. 3a, the energy and in Fig. 3b, number of scattering suffered by an emergent 
photon is shown. The higher energy photons are generated after relatively 
higher number of scatterings. The highest possible temperature is located at 
the centre of the disk $r_c$. The smooth energy contours like Fig. 3a can only be 
found for inclination independent cross-sectional view of the CENBOL. To an 
observer at a particular inclination angle, only a fraction of those photons 
reach the eye. Thus we get a rather noisier image.

\begin{figure}
\centering
\vbox{
\includegraphics[scale=0.6,angle=0,width=5.0cm]{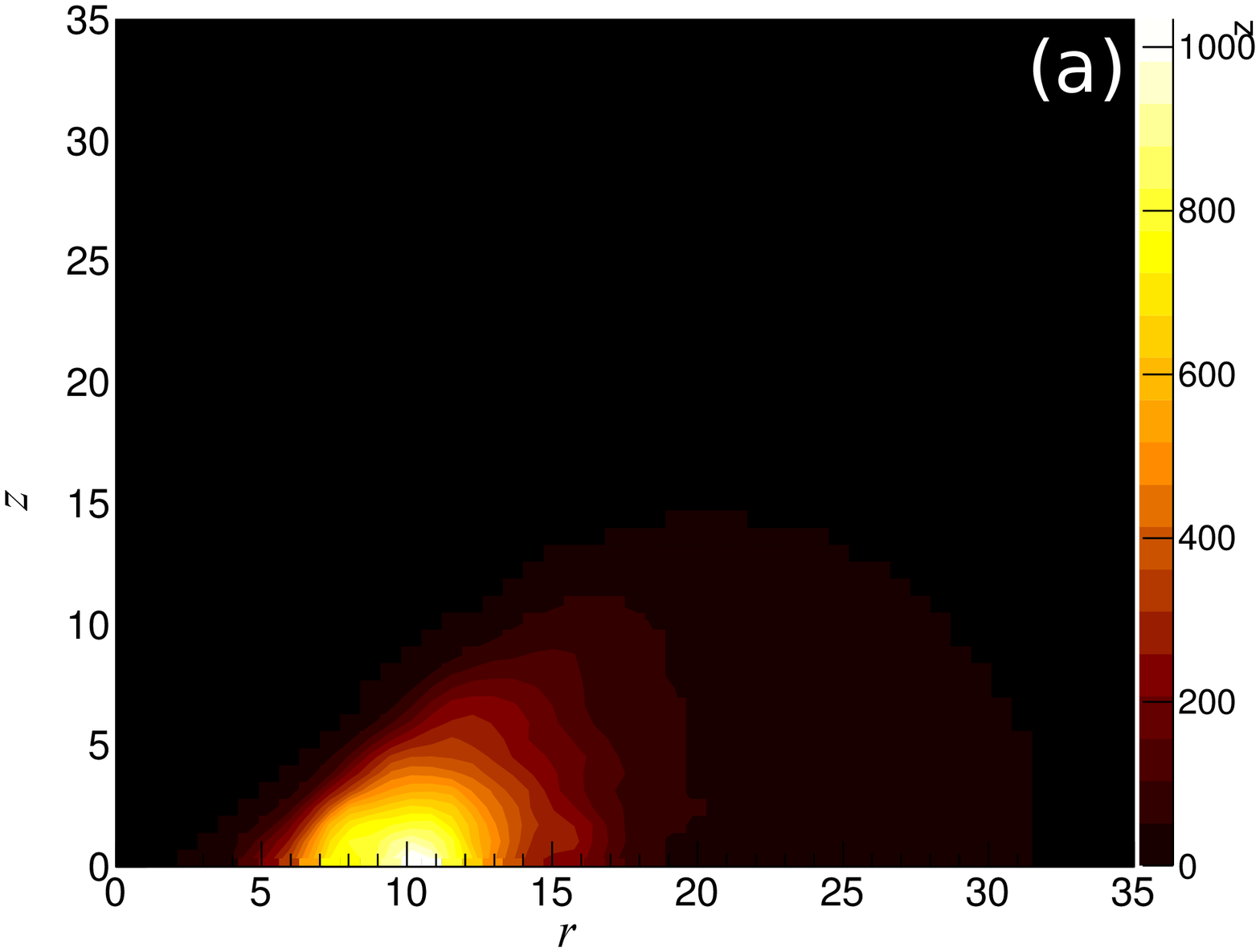}
\includegraphics[scale=0.6,angle=0,width=5.0cm]{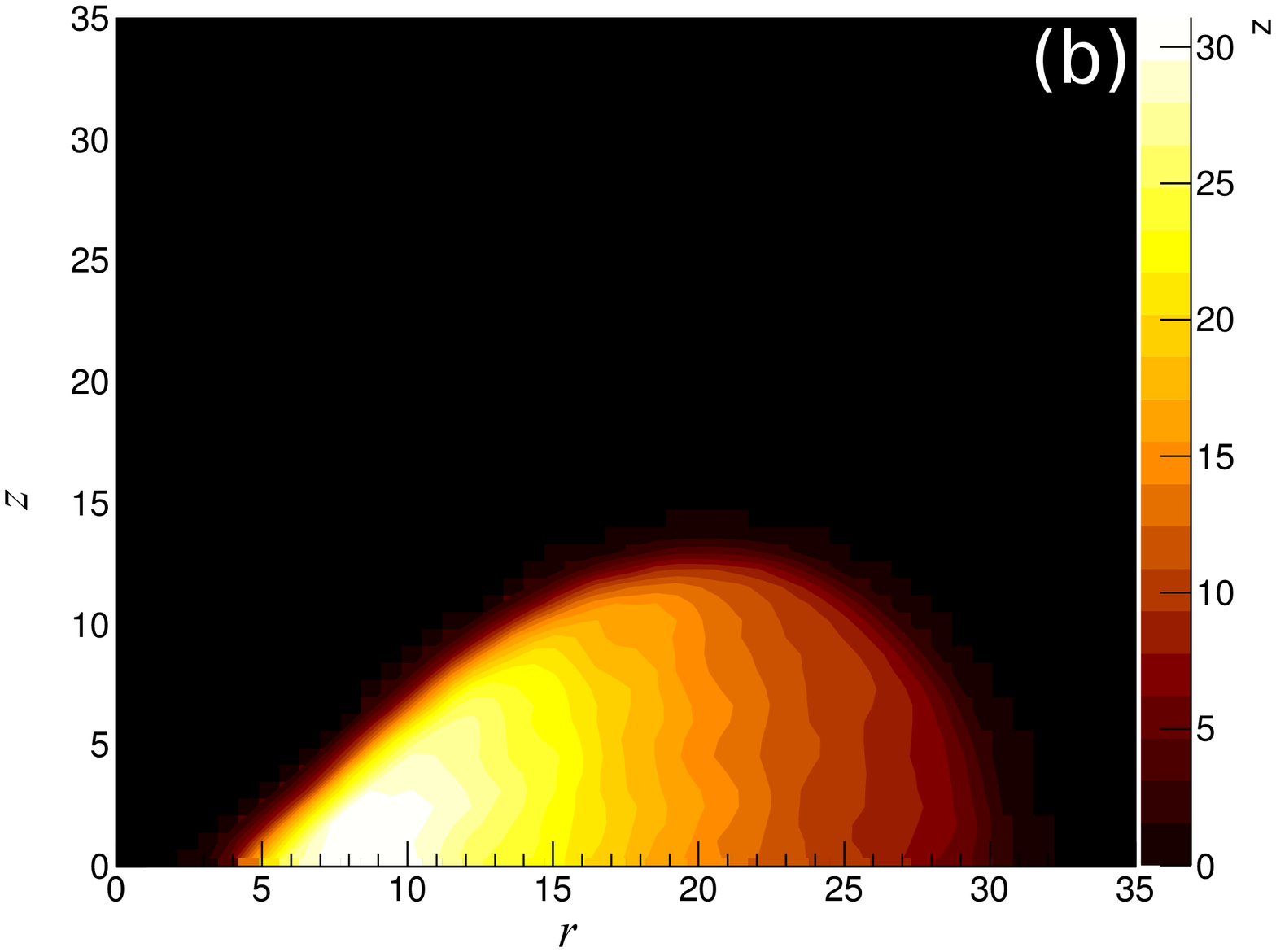}}\vskip 0.2cm
\caption{(a) Energy and (b) number of scattering suffered by emergent photons from the CENBOL. 
Photons which escape from the centre of CENBOL suffered maximum number of scattering. They generally
reach to highest energies due to Comptonization.} 
\end{figure}
      
\section{Space-Time Metrics used and the Image Generation Process}

In this Section, we describe our method of image generation process in details. 
For simplicity of the problem we choose the Schwarzschild geometry.  This describes 
the spacetime around a non-rotating, uncharged, spherically symmetric object. 

The metric components are assigned as (Weinberg, 1972), $g_{tt}=f(r)=(1-\frac{2}{r})$, 
$g_{rr}=h(r)=f(r)^{-1}$, $g_{\theta\theta}=-r^{2}$, $g_{\phi\phi}=-r^{2}\mathrm{sin}^{2}\theta$ and 
the inverse metric components as $g^{tt}=-f(r)^{-1}$,  $g^{rr}=h(r)^{-1}$, $g^{\theta\theta}=r^{-2}$,
 $g^{\phi\phi}=r^{-2}\mathrm{sin}^{-2}\theta$. Relation between curvature tensor 
and the Christoffel symbol follows
\begin{equation}
{\Gamma}_{\nu\lambda}^{\mu}= \frac{1}{2}g^{\mu\xi}\bigg[\frac{\partial g_{\xi\nu}}{\partial x^{\lambda}}
+\frac{\partial g_{\xi\lambda}}{\partial x^{\nu}}-\frac{\partial g_{\nu\lambda}}{\partial x^{\xi}}\bigg],
\end{equation}
where ${\Gamma}_{\nu\lambda}^{\mu}$ is the Christoffel symbols.
Non vanishing components of Christoffel symbols are
${\Gamma}_{rr}^{r}= \frac{1}{2h(r)}\frac{dh(r)}{dr}$, ${\Gamma}_{\theta\theta}^{r}=\frac{r}{h(r)}$,
${\Gamma}_{\phi\phi}^{r}=\frac{r^2\mathrm{sin}^2\theta}{h(r)}$,
${\Gamma}_{r\theta}^{\theta}={\Gamma}_{{\theta}r}^{\theta}={\Gamma}_{r\phi}^{\phi}
={\Gamma}_{{\phi}r}^{\phi}=\frac{1}{r}$,
${\Gamma}_{\phi\phi}^{\theta}=-\mathrm{sin}{\theta}\mathrm{cos}{\theta}$,
${\Gamma}_{\theta\phi}^{\phi}={\Gamma}_{\phi\theta}^{\phi}=\mathrm{cot}{\theta}$ and
${\Gamma}_{rt}^{t}={\Gamma}_{tr}^{t}=\frac{1}{2f(r)}\frac{df(r)}{dr}$ similar to
Weinberg (1972).

\subsection{Photon Trajectory Equations}
In four dimensions, the motion of free particles or photons are governed by the 
following equation,
\noindent
\begin{equation}
\frac{d^2x^{\mu}}{dp^2}+ {\Gamma}_{\nu\lambda}^{\mu}\frac{dx^{\nu}}{dp}\frac{dx^{\lambda}}{dp} = 0,
\end{equation}
with ${\mu} = [0,1,2,3]$; $x^{0} = t$, $x^{1} = r$, $x^{2} = \theta$ and $x^{3} = \phi$,
where $p$ is our affine parameter.

From Eqs $(1)$, $(2)$ and $(3)$ we obtain four coupled, second order differential 
equations. They are as follows:
\begin{equation}
\begin{aligned}
\frac{d^2t}{dp^2} + \frac{f'(r)}{f(r)}\bigg(\frac{dt}{dp}\bigg)\bigg(\frac{dr}{dp}\bigg)  =  0,\\
\frac{d^2r}{dp^2} + \frac{h'(r)}{2h(r)}\bigg(\frac{dr}{dp}\bigg)^2 - 
\frac{r}{h(r)}\bigg(\frac{d\theta}{dp}\bigg)^2 - \frac{r\mathrm{sin}^2{\theta}}{h(r)}\bigg(\frac{d\phi}{dp}\bigg)^2 
+ \frac{f'(r)}{2h(r)}\bigg(\frac{dt}{dp}\bigg)^2  =  0,\\
\frac{d^2\theta}{dp^2} + \frac{2}{r}\bigg(\frac{d\theta}{dp}\bigg)\bigg(\frac{dr}{dp}\bigg) 
- \mathrm{sin}{\theta}\mathrm{cos}{\theta}\bigg(\frac{d\phi}{dp}\bigg)^2 = 0 ~ \mathrm{and} \\
\frac{d^2\phi}{dp^2} + \frac{2}{r}\bigg(\frac{d\phi}{dp}\bigg)\bigg(\frac{dr}{dp}\bigg) 
+ 2\mathrm{cot}{\theta}\bigg(\frac{d\theta}{dp}\bigg)\bigg(\frac{d\phi}{dp}\bigg) = 0.
\end{aligned}
\end{equation}
We introduce energy $P_t=E=(1-\frac{2}{r})\frac{dt}{dp}$ and angular 
momentum $P_{\phi}=L=r^{2}\mathrm{sin}^{2}\theta\frac{d\phi}{dp}$ which are two physical
quantities for each photon (Chandrasekhar, 1983). Since the path of photons are 
independent of energy, we consider $P_t=E=1$. On the other hand, $P_{\phi}=L$ 
is a variable which essentially produces photons of various impact parameters  
as defined by $b=L/E$ (Luminet 1979).

Without the lose of generality, we can drop one equation by substituting 
the derivative from affine parameter $p$ to time $t$ co-ordinate. Thus we get
\begin{equation}
\begin{aligned}
\frac{d^2r}{dt^2} + \frac{3}{r(r-2)}\bigg(\frac{dr}{dt}\bigg)^2 - 
(r-2)\bigg(\frac{d\theta}{dt}\bigg)^2 - (r-2)r\mathrm{sin}^2{\theta}\bigg(\frac{d\phi}{dt}\bigg)^2 
+ \frac{r-2}{r^3} = 0,\\
\frac{d^2\theta}{dt^2} + \frac{2r-6}{r(r-2)}\bigg(\frac{d\theta}{dt}\bigg)\bigg(\frac{dr}{dt}\bigg) 
- \mathrm{sin}{\theta}\mathrm{cos}{\theta}\bigg(\frac{d\phi}{dt}\bigg)^2 = 0 ~\mathrm{and}\\
\frac{d^2\phi}{dt^2} +\frac{2r-6}{r(r-2)}\bigg(\frac{d\theta}{dt}\bigg)\bigg(\frac{dr}{dt}\bigg)
 + 2\mathrm{cot}{\theta}\bigg(\frac{d\theta}{dt}\bigg)\bigg(\frac{d\phi}{dt}\bigg) = 0.
\end{aligned}
\label{eq:xdef}
\end{equation}
Finally, we have three second order coupled differential equations which governs 
the path of a photon in this curved geometry.

\subsection{Velocity Components: Tetrad formalism}

The formalism of Tetrad is used to connect various coordinates. The nature of 
Tetrad frame is locally inertial (Park, 2006). By virtue of this, it connects 
physical quantities in curved spacetime to a flat spacetime. Fixed and co-moving 
Tetrad are of most importance. However, here, we shall concentrate our studies 
for fixed Tetrad. Let us take an orthonormal fixed Tetrad (for our observer) 
with basis vector ${e_{\hat{\mu}}} = \partial/\partial {\hat x^{\mu}}$. The 
connection to another coordinate with unit vector ${e_{\mu}} = \partial/\partial x^{\mu}$ 
is established via following relations (Weinberg 1972):
\begin{equation}
\begin{aligned}
\displaystyle \frac{\partial}{\partial \hat{t}} = \frac{1}{f(r)^{1/2}}\frac{\partial}{\partial t},\\ 
\displaystyle \frac{\partial}{\partial \hat{r}} = f(r)^{1/2}\frac{\partial}{\partial r},\\
\displaystyle \frac{\partial}{\partial \hat{{\theta}}} = \frac{1}{r}\frac{\partial}{\partial \theta} ~\mathrm{and} ~\\
\displaystyle \frac{\partial}{\partial \hat{{\phi}}} = \frac{1}{r\mathrm{sin}\theta}\frac{\partial}{\partial \phi}.\\
\end{aligned}
\label{eq:xdef}
\end{equation}

These transformations allow us to connect physical quantities from the curved to 
the flat spacetime. From above equations one can deduce three spatial velocity 
components which can be expressed as,
\begin{equation}
\begin{aligned}
v^{\hat{r}}=\frac{d\hat{r}}{dt}=\frac{r}{(r-2)}\frac{dr}{dt},~v^{\hat{\theta}}=\frac{d\hat{\theta}}{dt}=\frac{r\sqrt{r}}{\sqrt{(r-2)}}\frac{d\theta}{dt}\\ \mathrm{and} ~
v^{\hat{\phi}}=\frac{d\hat{\phi}}{dt}=\frac{r\sqrt{r}\mathrm{sin}{\theta}}{\sqrt{(r-2)}}\frac{d\theta}{dt}.
\end{aligned}
\label{eq:xdef}
\end{equation}

\subsection{Solving Geodesic Equations}

We solve these three differential equations using fourth order Runge - Kutta method where
initial velocity components are supplied in Cartesian co-ordinate. The unitary
transformation matrix between $(v^{\hat{x}}, v^{\hat{y}},v^{\hat{z}})$ and $(v^{\hat{r}},
v^{\hat{\theta}},v^{\hat{\phi}})$ is given below,

\begin{equation}
\centering
\left(
\begin{array}{c}
v^{\hat{r}}\\
v^{\hat{\theta}}\\
v^{\hat{\phi}}
\end{array}
\right)=
\left(
\begin{array}{ccc}
 \mathrm{sin}{\theta}\mathrm{cos}{\phi}&\mathrm{sin}{\theta}\mathrm{sin}{\phi}&\mathrm{cos}{\theta}\\
  \mathrm{cos}{\theta}\mathrm{cos}{\phi}&\mathrm{cos}{\theta}\mathrm{sin}{\phi}&-\mathrm{sin}{\theta}\\                 
  -\mathrm{sin}{\theta}&\mathrm{cos}{\phi}&0\\
\end{array}
\right)
\left(
\begin{array}{c}
v^{\hat{x}}\\
v^{\hat{y}}\\
v^{\hat{z}}
\end{array}
\right).
\end{equation}
The initial positions ($x^{\nu}_{\circ}$, where $\nu=1, 2, 3$) and velocities
($\left[\frac{d\hat{x}^{\nu}}{dt}\right]_{\circ}$, with $\nu=1, 2, 3$) of emergent
photons are supplied from the Monte-Carlo Comptonization code. To obtain images 
for a particular inclination angle, we combine all the photons obtained from various 
azimuthal angles to increase the intensity of the image. The rotational symmetry in 
Schwarzschild geometry allows us to do this superposition without a loss of generality.

\subsection{Redshifts of Photons}

To generate spectrum and the image one must introduce the redshift. In this scenario,
there are two parts of the redshift. One part comes from the Doppler effect caused by rotational
motion of the disk matter and the other part is the gravitational redshift caused by the 
black hole. So, the net redshift can be written as,
\begin{equation}
1+z= r_{grav}r_{rot},
\end{equation}
where, $r_{grav}$ and $r_{rot}$ are gravitational and rotational contributions to redshift
respectively. As a whole, one can express redshift as,
\begin{equation}
1+z= \frac{E_{em}}{E_{obs}}=\frac{(P_{\alpha}u^{\alpha})^{em}}{(P_{\alpha}u^{\alpha})^{obs}},
\end{equation}
where, $E_{em}$ and $E_{obs}$ are the energy of emitted and observed photons respectively. Now,
taking an inner product of 4-momentum and 4-velocity gives the energy of emitted photon as
\begin{equation}
 E_{em}=P_tu^t+P_{\phi}u^{\phi}=P_tu^t\bigg(1+\Omega_{\phi}\frac{P_{\phi}}{P_t}\bigg),
\end{equation}
where $\Omega_{\phi}=\frac{u^{\phi}}{u^t}$ (for Keplerian disk $\Omega_{\phi}=(1/r^3)^{1/2})$
and for CENBOL region $\Omega_{\phi}=c\lambda^{n-2}$). From earlier definition
$\frac{P_{\phi}}{P_t}=\frac{L}{E}$ and known as the projection of
impact parameter along the z-axis. As in Luminet (1979), we consider the final form of redshift as,
\begin{equation}
1+z = (1-3/r)^{-1/2}(1+\Omega_{\phi}b\mathrm{sin}\theta_{\circ}\mathrm{sin}{\alpha}),
\end{equation}
where, $(1-3/r)^{-1/2}=r_{grav}$ and $(1+\Omega_{\phi}b\mathrm{sin}\theta_{\circ}\mathrm{sin}{\alpha})
=r_{rot}$. The relation between the observed flux and the emitted flux is given by,
\begin{equation}
F_{k}^{obs}= \frac{F_{k}^{disk}}{(1+z)^4}.
\end{equation}
Fourth power in the redshift factor comes due to energy loss of photons, gravitational
time dilation and relativistic correction of detector solid angle. Similarly, the
observed temperature will differ from the source temperature by,
\begin{equation}
T_{k}^{obs} = \bigg(\frac{F_{k}^{obs}}{\sigma}\bigg)^{1/4}.
\end{equation}
For a Keplerian disk, the maximum observed flux is located at $r(F_{k}^{max})=9.77$
with $F_{k}^{max}\sim1.763\times10^{-4}~F_c(\dot{m_{d}})$ (all accretion 
rates are measured in units of $\dot{m_{d}} =\frac{\dot{M}}{\dot{M}_{edd}}$).

\begin{figure}
\centering
\includegraphics[width=.3\textwidth,angle=0]{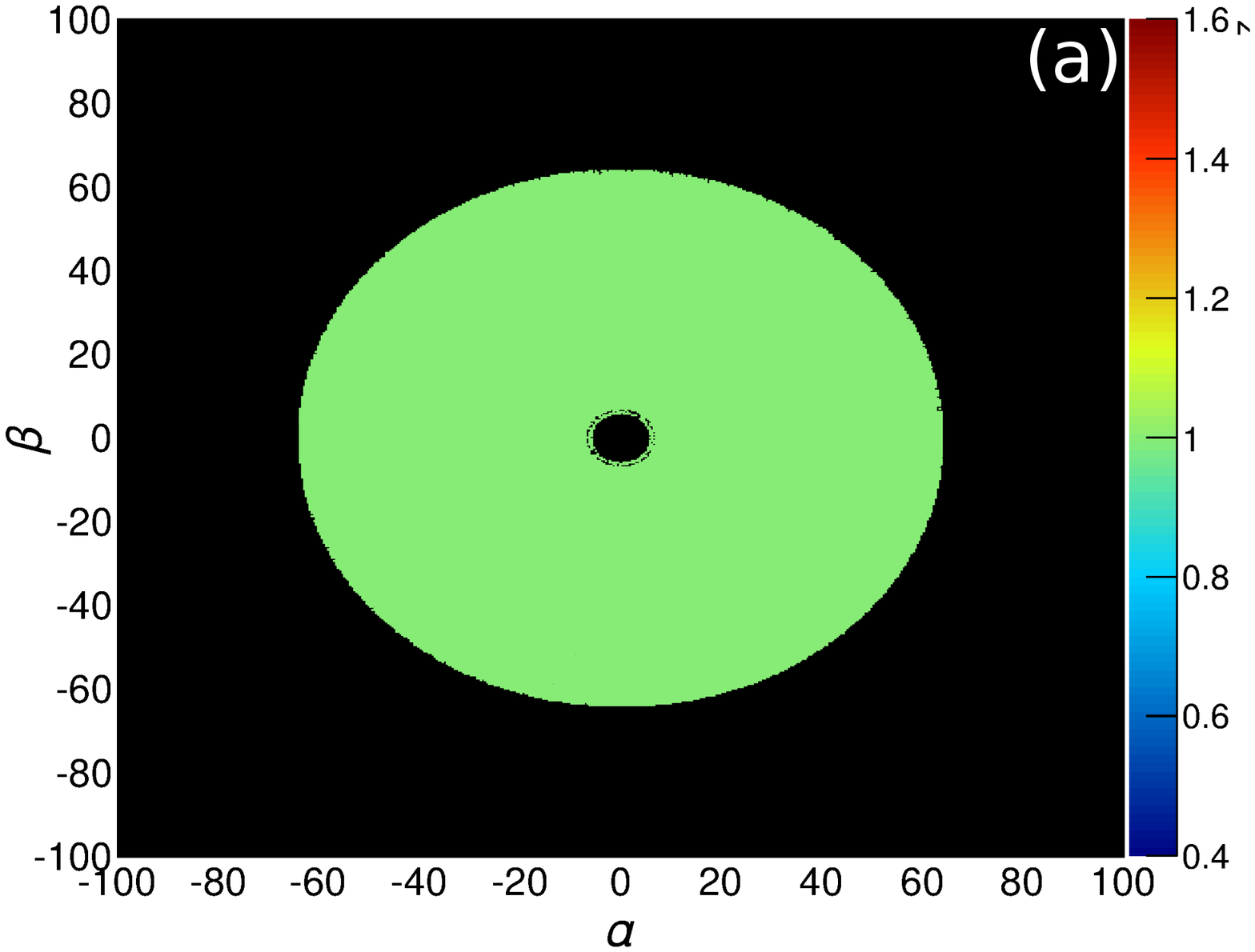}
\includegraphics[width=.3\textwidth,angle=0]{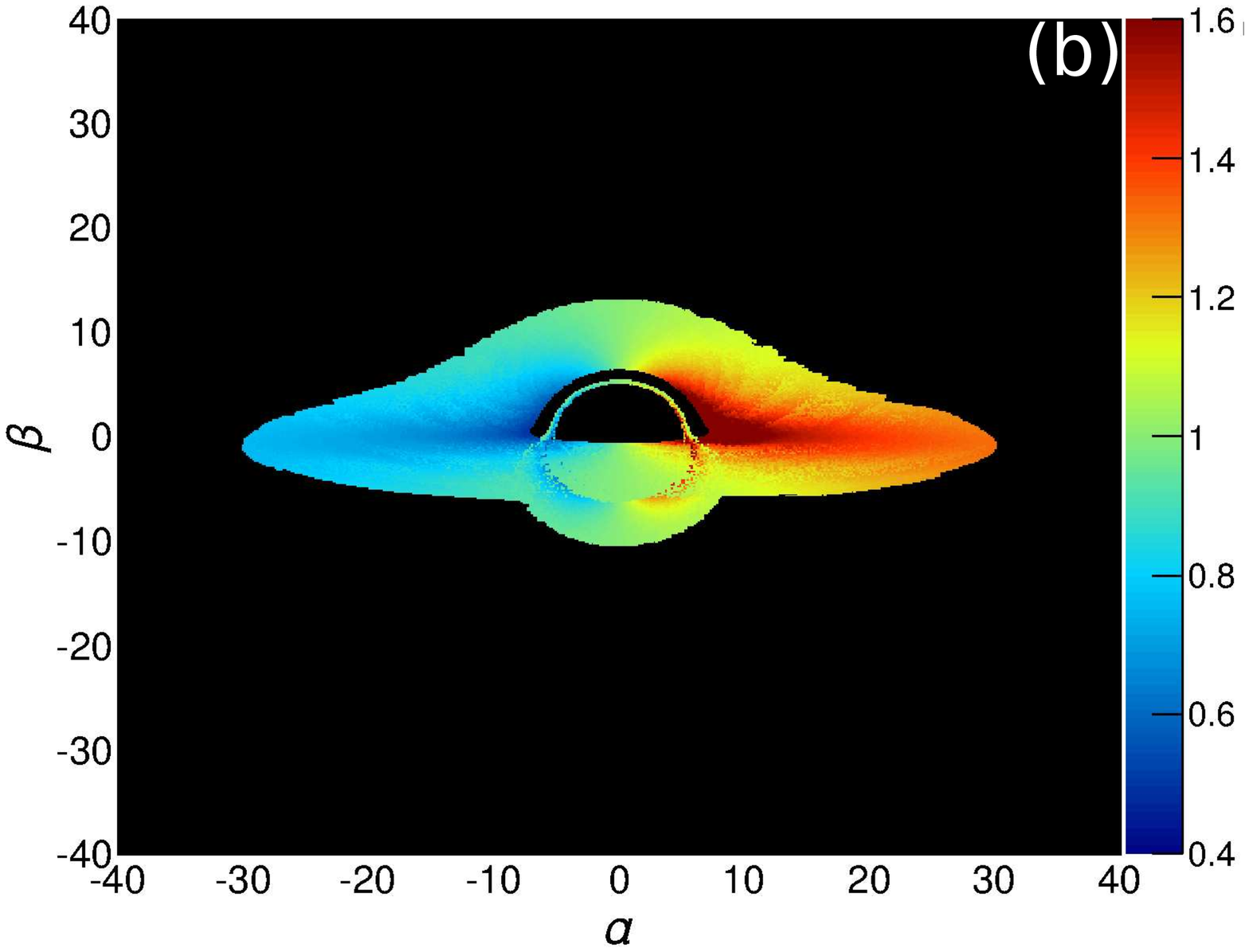}
\includegraphics[width=.3\textwidth,angle=0]{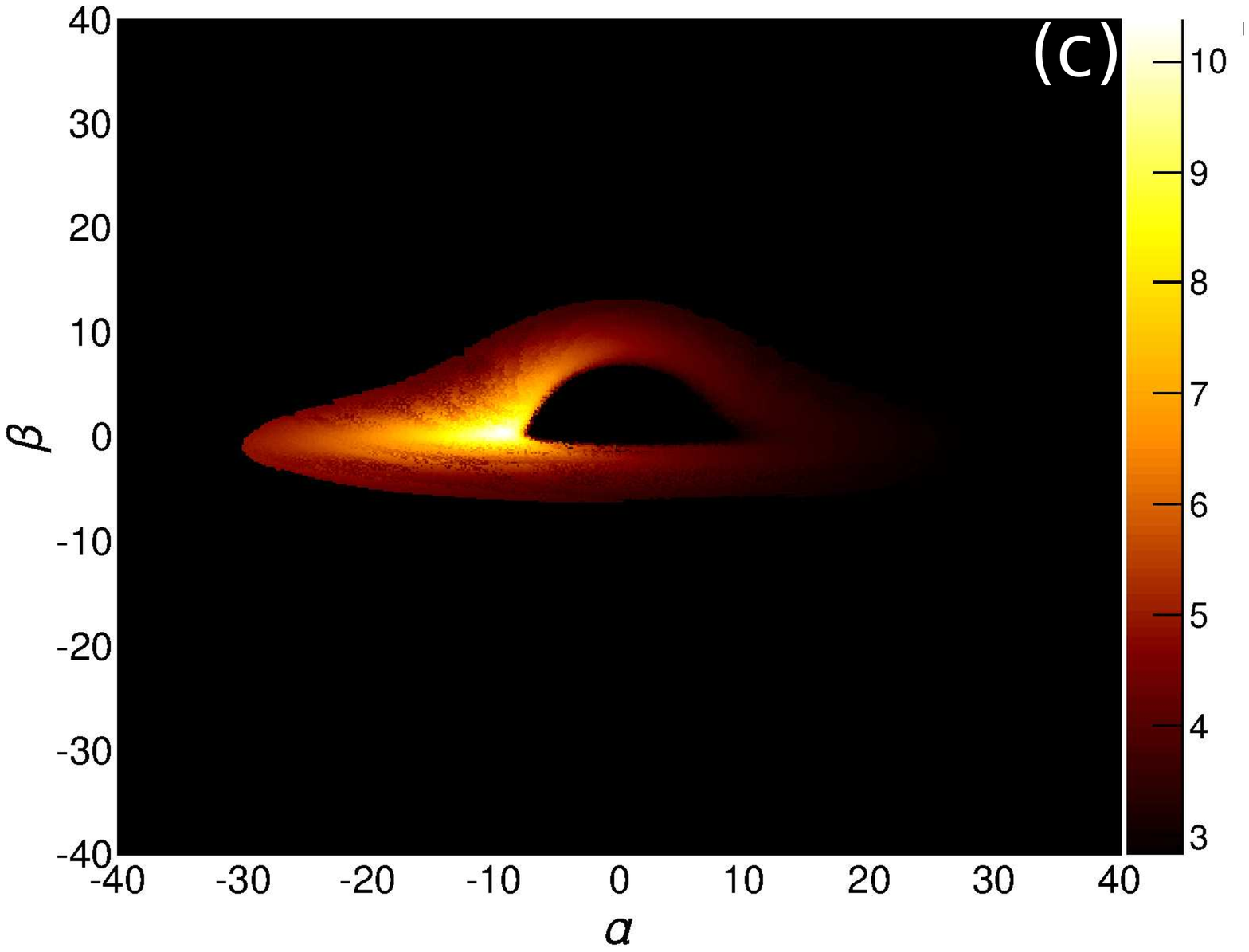}
\caption{Doppler tomography of Keplerian disk having
inclination angles (a) $0.01^{\circ}$ and (b) $82^{\circ}$ with outer edge ($r_{out}$)
at $60.0$ and $30.0$, respectively. Exact pole on view will give a complete
zero Doppler effect disk. Colors represent contours of $(1+r_{rot})$. Image of Keplerian
disk with $r_{out}=30.0$ at an inclination angle $82^{\circ}$ is shown in (c).
The color bar represents normalized temperature of the Keplerian disk.
The inner edge ($r_{in}$) of the Keplerian disk is at $6.0$ in all these three Figures.}
\end{figure}

The Doppler effect increases as we increase the inclination angle. In case of 
$0^{\circ}$ inclination, there will not be any line splitting due to Doppler 
effect (Fabian et al., 2000). We show Doppler tomography for $0.01^{\circ}$ (Fig. 4a) 
inclination angle where the effect of differential rotation of the accretion disk 
is negligible and the observed spectra remain similar to the emitted spectra. In Fig. 4b,
we see the effects of rotation for $82^{\circ}$ inclination, where the iso-radial 
curves are deformed and observed spectra differ from what the disk is emitting. 
We consider all the photons in generating tomographic Figures. The image of the 
Keplerian disk is shown in Fig. 4c for an inclination angle of $82^{\circ}$. 
The color-bar represents a normalized observed temperature of Keplerian disk with 
the inner edge at $6.0$ and outer edge at $30.0$. We consider first order 
photons (i.e., photons which reach us directly without a complete rotation around 
the black hole) to construct this image. Because gravitational redshifts of photons 
which arrive at the observer after orbiting the black hole at least once or more is high,
the contribution of those photons in a realistic image will be very low. 

A picture similar to Fig. 4c has previously been presented in Fukue \& Yokoyama (1988) using
the semi-analytical transfer function. The null trajectory solution (Eqn. 11) which we use, 
gives us the freedom to use three dimensional geometry and ability to gather information of 
physical quantities of that photon at any point in their path. The time of arrival of photons 
to an observer can be evaluated by this process. Using Monte Carlo technique, we produce 
an image, more realistic than the earlier images using semi-analytical transfer function methods.

\section{Results and Discussions}

\subsection{Spectra}

Now that we have demonstrated our ability to image a standard disk, we now carry 
out the Monte-Carlo simulation including Comptonization and consider a TCAF configuration 
with a CENBOL and a truncated Keplerian disk as described in \S 2. We keep our detectors 
at $50$, just outside the TCAF system to get the simulated source spectrum, i.e., the spectrum that is just 
coming out of the source. For the generation of the image, the observer 
is placed at $100$. In Fig. 5(a), the hardening of the 
output spectrum with the increase of the central temperature of the CENBOL is shown. Here, 
the disk rate ($\dot{m_{d}}$) is kept fixed at $1.0$ Eddington rate. For spectral study, 
We define the energy spectral index $\alpha$ to be $I(E) \propto E^{-\alpha}$. We note 
that $\alpha$ decreases, i.e., the spectrum hardens with the increase in the central 
temperature ($T_c$) of CENBOL. The spectrum of the injected soft photons for accretion 
rates of $\dot{m_{d}}=0.00001~\mathrm{(solid-black)}$, $ ~\dot{m_{d}}=0.001~\mathrm{(dashed-red)}$ 
and $~\dot{m_{d}}=0.1~\mathrm{(dot-dashed-green)}$ are shown in Fig. 5b. The
 corresponding emergent spectra keeping the central temperature fixed at 
200keV is shown in Fig. 5c. In Fig. 5d, the direction dependent source 
spectrum is shown. Photons are placed in four equal sized bins from 0 to 90 degrees
($0^{\circ}-22.5^{\circ}$, $22.5^{\circ}-45^{\circ}$, $45^{\circ}-67.5^{\circ}$ and $67.5^{\circ}-90^{\circ}$
respectively.) The spectrum is clearly getting harder for higher inclination angles. Variation of 
$\alpha$ with $T_c$ and $\dot{m_{d}}$ is presented also in Table 1. Also, $\alpha$ 
increases with increasing $\dot{m_{d}}$. The result is consistent with those of 
Chakrabarti \& Titarchuk (1995) carried out in the pseudo-Newtonian geometry.
\begin{figure}
\vskip 1cm
\centering
\vbox{\includegraphics[width=13.0cm,height=8cm]{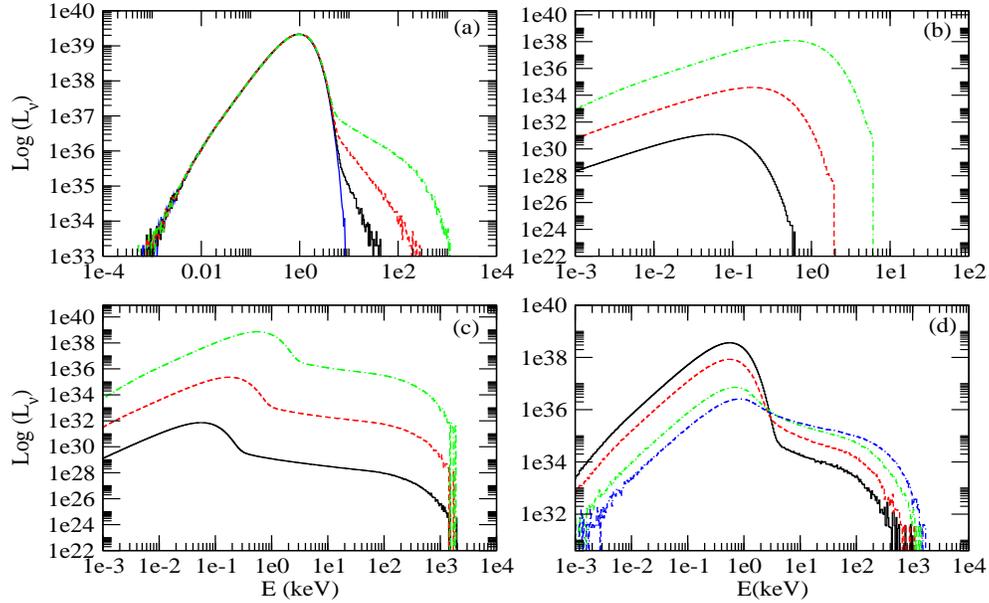}}\vskip 0.5cm
\caption{Variation of emergent spectrum with respect to (a) central temperature of CENBOL
keeping $\dot{m_{d}}=1.0$ as fixed. The injected spectrum is shown by solid-blue line. The dashed-black,
dot-dashed-red and double-dash-dot-green spectra correspond to the central temperatures $50$, $100$ and
 $140$ keV, respectively. (b) Variation of injected seed photons from truncated disks
of accretion rates $\dot{m_{d}}=0.00001~\mathrm{(solid-black)}$, $ ~\dot{m_{d}}=0.001~\mathrm{(dashed-red)}$, 
$~\dot{m_{d}}=0.1~\mathrm{(dot-dashed-green)}$.
The central temperature is $200$ keV. (c) Emergent spectra with the seed photons as in (b).
(d) Direction dependent emergent spectra for $~\dot{m_{d}}=0.1$.  
The solid-black, dash-red, dot-dash-green and dot-dash-dash-blue curves represent
inclination angles bins of $0^{\circ}-22.5^{\circ}$, $22.5^{\circ}-45^{\circ}$, $45^{\circ}-67.5^{\circ}$ and $67.5^{\circ}-90^{\circ}$
respectively.}
\end{figure}

\begin{table}
\centering
\caption{$\alpha$ variation with $T_c$ \& $\dot{m_{d}}$}
\begin{tabular}{cccc}
  \toprule[1.5pt]
Figure & $\dot{m_{d}}$ & $T_c (keV)$ & $\alpha$  \\
  \hline
5a (solid − black)         &  1.0     & 50.0   & 2.15  \\ 
5a (dashed − red)          &  1.0     & 100.0  & 1.81  \\ 
5a (dot − dashed − green)  &  1.0     & 140.0  & 1.09  \\ \hline
5c (solid − black)         &  0.1     & 200.0  & 0.58  \\ 
5c (dashed − red)          &  0.001   & 200.0  & 0.55  \\
5c (dot − dashed − green)  & 0.00001  & 200.0  & 0.54  \\ 
  \bottomrule[1.5pt]
\end{tabular}
\end{table}

In Fig. 6, we show the variation of spectrum as seen by the observers sitting at three
different inclination angles. The spectrum is getting harder as the inclination angle 
increases. This is clearly seen by the increase in the number of power-law photons for 
higher $\theta_{obs}$. For a photon to reach an observer at a higher $\theta_{obs}$ 
it must pass through the CENBOL region and thus the probability of scattering increases. 
This makes the spectrum at higher $\theta_{obs}$ harder. We repeat the process for three
different disk accretion rates and observe softening of the spectrum with the increase in
$\dot{m_{d}}$. 

\begin{figure}
\vskip 1cm
\centering
\vbox{
\includegraphics[width=13.0cm,height=4.0cm]{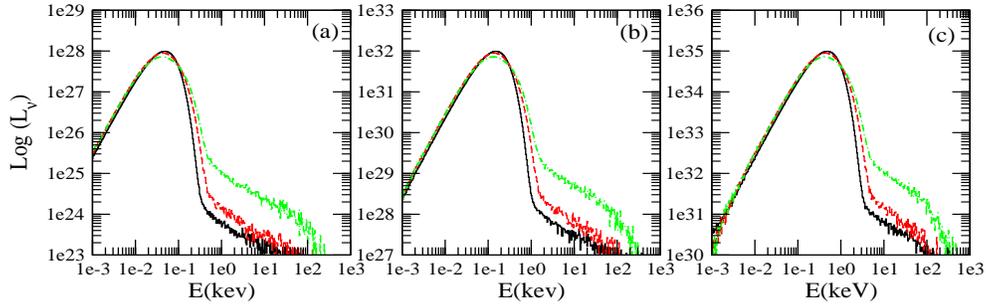}}\vskip 0.5cm
\caption{Variation of simulated observed spectrum with inclination angles $10^{\circ}$ (solid-black),
$50^{\circ}$ (dot-red) and $80^{\circ}$ (dash-green). Figs. (a), (b) and (c) corresponds to
$\dot{m_{d}}=10^{-5}$, $10^{-3}$ and $10^{-1}$, respectively. Hardening of spectrum with
the increase in inclination angle is clearly seen.}
\end{figure}

Fig. 7 shows the variation  of the ratio of $N_{pl}$ (the number of Comptonized 
photons) and $N_{b}$ (the number of black body photons) with inclination angles 
for accretion rates $\dot{m}_d=10^{-1}$ (solid black), $10^{-3}$ (dashed red) 
and $10^{-5}$ (dot-dashed-green) respectively. This ratio increases with decreasing 
accretion rate for a fixed inclination angle and increases with increasing 
inclination angle for a fixed accretion rate. The increment remains small for 
$10^{\circ}-50^{\circ}$ inclination. But, for $\theta_{obs} \gsim 60^{\circ}$, the
power-law photon number increases rapidly. The edge on view of CENBOL provides 
higher number of Comptonized photons. Thus, we can generally expect smaller spectral 
index or harder states  and more spectral variabilities for higher inclination 
angle objects.

The black-body part of high inclination angle bin ($67.5^{\circ}-90^{\circ}$) of the 
{\it simulated source spectrum} in Fig. 5d is much flatter than the low inclination angle case. 
But, in the {\it simulated observed spectrum} (Fig. 6) obtained at $100$, the 
black-body part is much higher. Due to the shadowing effect of CENBOL, less number of 
black-body photons reach when we placed detectors just outside the system at $50$ (Fig. 5d). 
The effect decreases with increasing observer distance. However, if the size of the CENBOL 
is large, the shadowing effect will also increase. This is a clear effect of photon 
bending where focussed photons which marginally escaped the CENBOL region arrive to observer 
at higher inclination angle.

\begin{figure}
\vskip 1cm
\centering
\vbox{
\includegraphics[width=8.0cm,height=6.0cm]{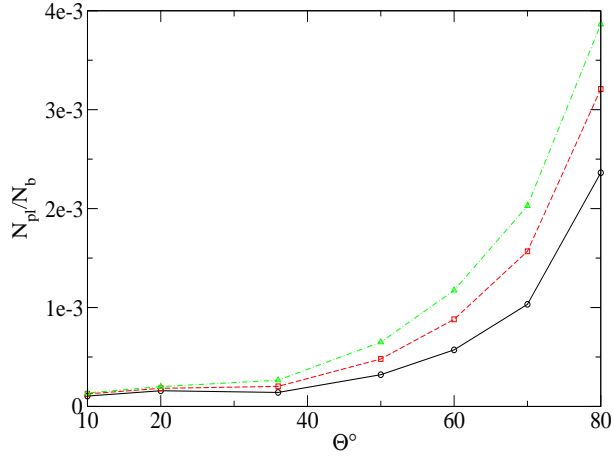}}\vskip 0.5cm
\caption{Variation of photon counts with inclination angle for accretion rates 
$\dot{m_{d}}=10^{-1}$~(solid-black), $10^{-3}$~(dashed-red) and $10^{-5}~(dot-dashed-green)$ 
is shown here. $N_{b}$ and $N_{pl}$ are black-body and power-law photon counts, 
respectively. As the inclination angle increases, the number of power-law photon 
increases. The black-body cut-offs for $\dot{m_{d}}=10^{-1}$, $10^{-3}$ and $10^{-5}$ 
accretion rates are considered as $3.5$, $1.0$ and $0.3$ keV respectively.}
\end{figure}

\begin{figure}
\vskip 1cm
\centering
\vbox{
\includegraphics[width=13.0cm,height=4.0cm]{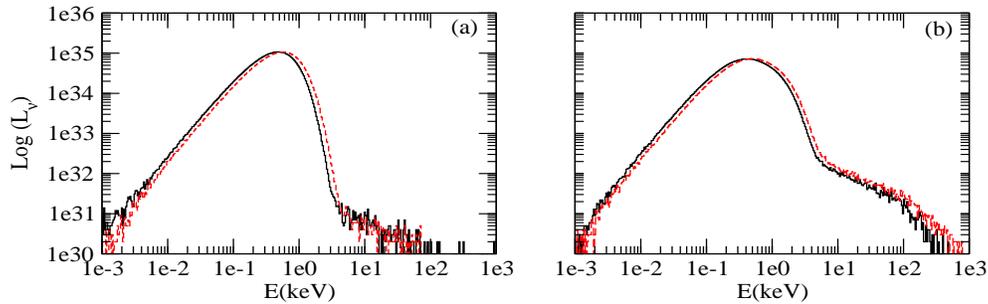}}\vskip 0.5cm
\caption{Variation of spectrum with and without gravitational bending for accretion 
rate $\dot{m_{d}}=0.1~$ is shown. In Fig. (a) and (b) inclination angles are $10^{\circ}$ 
and $80^{\circ}$, respectively. The solid-black line represents the spectra where gravitational 
bending is included. The dashed-red line represents where we considered straight line path
for photons.}
\end{figure}

We have shown simulated spectra as seen by an observer placed at $r=100$ for 
different accretion rates and inclination angles. We repeat the
simulation by considering that the photons follow a straight line path as in a 
Newtonian space time; i.e., after excluding the effect of gravitational bending 
of light. We added Doppler effect because its outcome will still be present in 
the Newtonian regime. However, gravitational redshift is not included. We show 
the comparative spectra of two inclination angles having the same accretion rates 
$\dot{m_{d}}=0.1~$ in Fig. 8. Due to the gravitational red-shift, the whole 
spectra for both cases shifted towards a lower energy range. Since the effect 
of gravitational redshift depends only on the mass of black hole, even after 
increasing inclination angle the difference between the two spectra remains the same.  

\subsection{Images}
Image construction requires a one to one connection between an emitted photon at 
the source and an observed photon in the observer plane. This is established 
using geodesic relations. Photons scatter all around but the observer is generally localized
($r_{obs},~\theta_{obs} ~\mathrm{and}~\phi_{obs}$). Exploiting the symmetry of
 the problem, we integrate over $\phi_{obs}$ so as to merge observers with different 
$\phi_{obs}$. Integration over elevation angle $\theta_{obs}$ cannot be done. The 
distribution in $\theta_{obs}$ enable us to compare spectra and imaging at various 
inclination angles. A pole on view of the disk gives only the isoradial circles and the
spectrum has no significant deviation from the source spectrum. The deviation of 
frequency and distortion in the isoradial curves became noticeable after 
$\theta_{obs}>30^{\circ}$ (Luminet, 1979).

\begin{figure}
\centering
\vbox{
    \includegraphics[width=1.0\textwidth,angle=0]{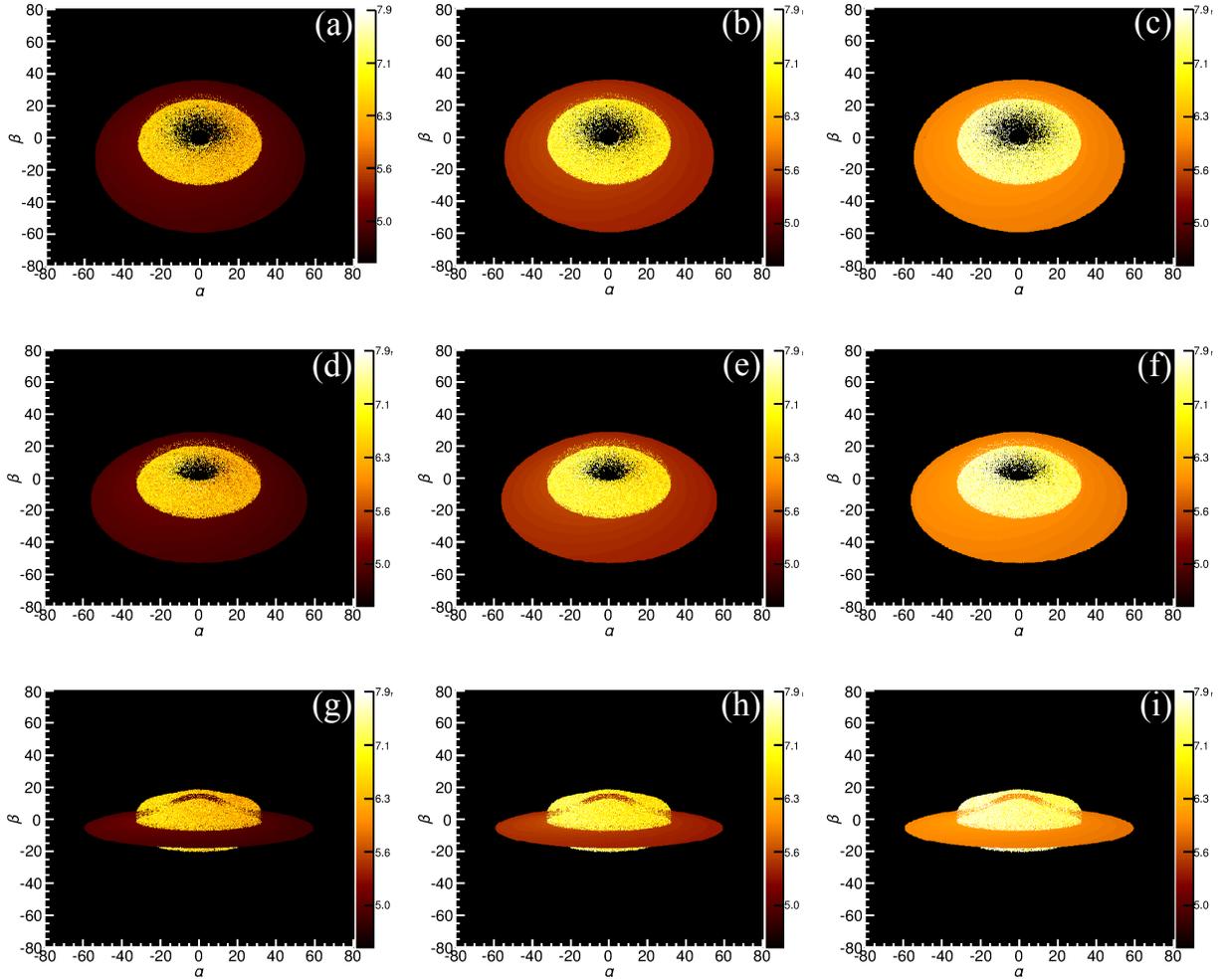}}\vskip 0.5cm
  \caption{Images seen by observers from (top panel) $36^{\circ}$, $50^{\circ}$ (middle panel), 
$80^{\circ}$ (bottom panel). Disk rates in the first, second and last column are $\dot{m_{d}}=0.00001$, 
$ ~\dot{m_{d}}=0.001$ and $~\dot{m_{d}}=0.1$ respectively. Color-bar is plotted in the 
$log(T_{obs})$ scale with $10^{4.5}-10^{7.9}~$ Kelvin is the range fixed for each Figure.}
\end{figure}

The image of CENBOL with a Keplerian disk around is realistic as the spectra 
of the combined flow explains the observations most successfully. In Fig. 9, we present 
images of TCAF at various inclination angles. We used all the photons which appeared 
at a given inclination angle in constructing these images. Here, we demonstrate our 
results at three inclination angles with different accretion rates. As the accretion 
rate is increased, the disk temperature increases. The peak energy of black-body 
shifts towards higher value. The intensity also increases. So, the disk color becomes 
brighter emitting higher frequency radiation with accretion rates. We have considered the temperature range 
such that all the photons from the disk and CENBOL can be shown together in a single 
plot where temperature contours of disk are uniformly binned. The band structure on the 
Keplerian disk is due to this binning. With increasing disk rate, the minimum energy of 
Comptonized photon increases. These photons are mostly generated on the outer boundary
of CENBOL. Color variation of the CENBOL region in Fig. 9 (from left to right accretion 
rate increases) is caused mostly due to the photons that are coming from the outer 
regions of the CENBOL. For high inclination ($\theta_{obs}=80^{\circ}$), Fig. 9g 
shows near disappearance of many Keplerian disk photons which were coming from 
the other side of the disk for a low disk accretion rate ($\dot{m_{d}}=0.00001$).
As the accretion rate goes down, the percentage of scattered photons increases. Since 
the photons from the opposite side (of the observer) which are coming towards the 
observer at an higher inclination angle must pass through the CENBOL region, they 
are more scattered. This interesting effect due to focusing could also be tested 
observationally in future, for high inclination sources. Interestingly, with an
increase in the disk rate, the number of photons passing through the CENBOL increases 
and the CENBOL appears to be splitted into two halves especially at high inclination 
angle. Note that the number of photons from the inner edge of the CENBOL is
very few and they are either sucked into the black hole or are absorbed. Note that the 
inner edge of CENBOL is noisy especially for lower accretion rates, since very few 
photons are intercepted by the CENBOL in the first place and even fewer 
are emitted from this region. This noisy behavior is inherent to Monte-Carlo simulations
and would go away with a larger sample of injected photons and also after smoothening 
out over finite detector pixels. The occasional color fluctuations of photons which 
appear to emerge side by side are due to the fact that they are originated 
from two different optical depths which may have different temperatures, 
gravitational redshifts and energy transfer through Comptonization. This is to 
be contrasted with the smooth image of a spherical Comptonizing corona having 
a constant temperature and optical depth as obtained by Schnittman \& Krolik (2009).

There are other effects in our image which need some attention.
The high energy photons (yellow) which are visible on the top of the distant
half of the Keplerian disk in Fig. 9(a-f) are simply due to warped nature of the
CENBOL. Just as in Fig. 4b where distant side of the Keplerian disk appeared to be warped,
the distant side of CENBOL here also appears to be warped with the moderately warped
Keplerian disk in the background. It is to be noted that the CENBOL emits only
those photons intercepted from the Keplerian disk and are much fewer in
number and thus the more intense, low temperature, Keplerian
disk can still be seen peeping through the CENBOL, specially when the optical depth is low.
For high inclination angle, only some of these Keplerian
photons are scattering by relatively tenuous CENBOL and thus the warped Keplerian disk
appears to be cutting the CENBOL into two parts. The front side of CENBOL is also
cut into two halves mainly because the outer edge of the Keplerian disk is chosen to be
at $50GM/c^2$.  A larger outer edge or a lower optical depth of the CENBOL would have removed
the lower half of the CENBOL completely.

\begin{figure}
\centering
\vbox{
    \includegraphics[width=0.31\textwidth,angle=0]{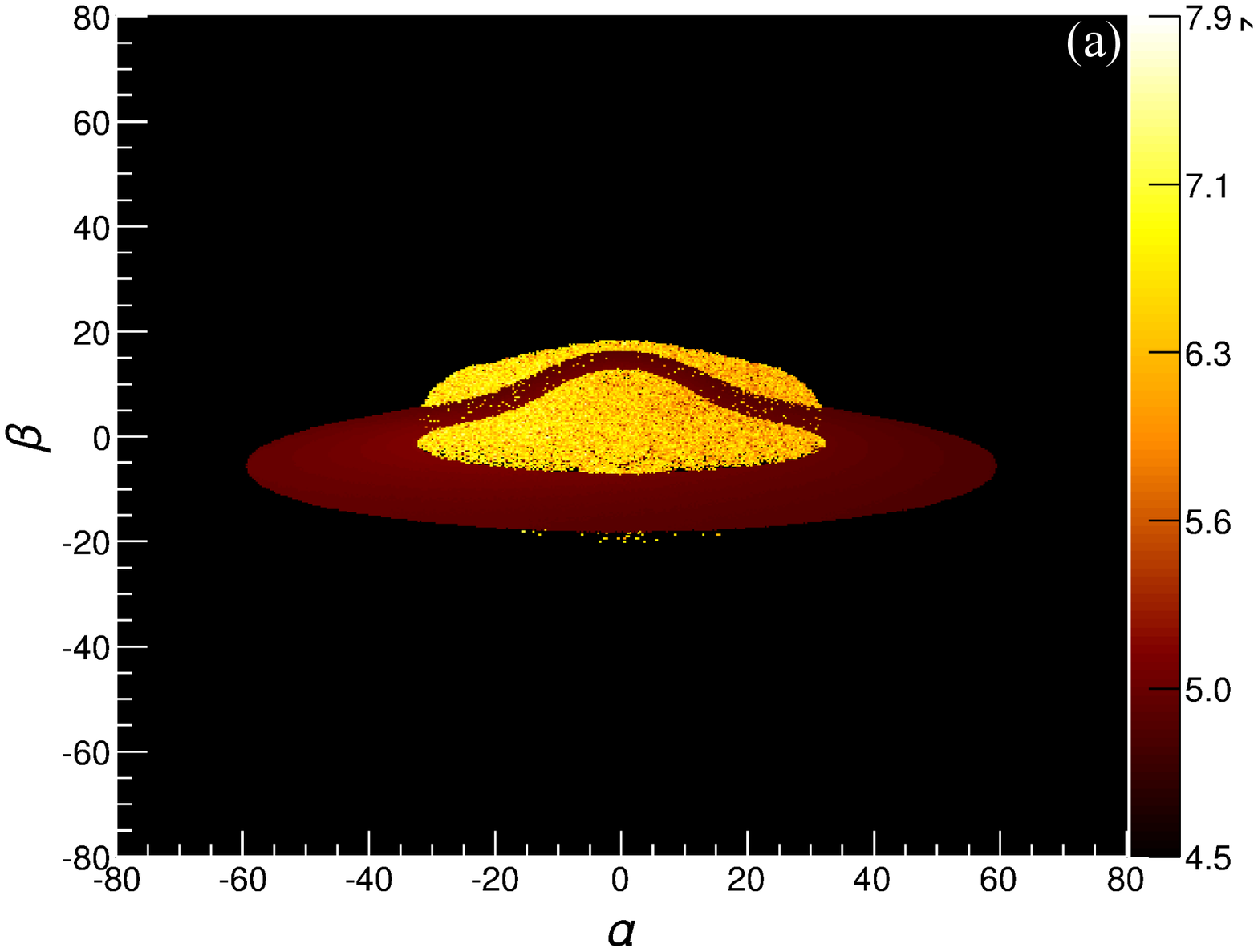}
    \includegraphics[width=0.31\textwidth,angle=0]{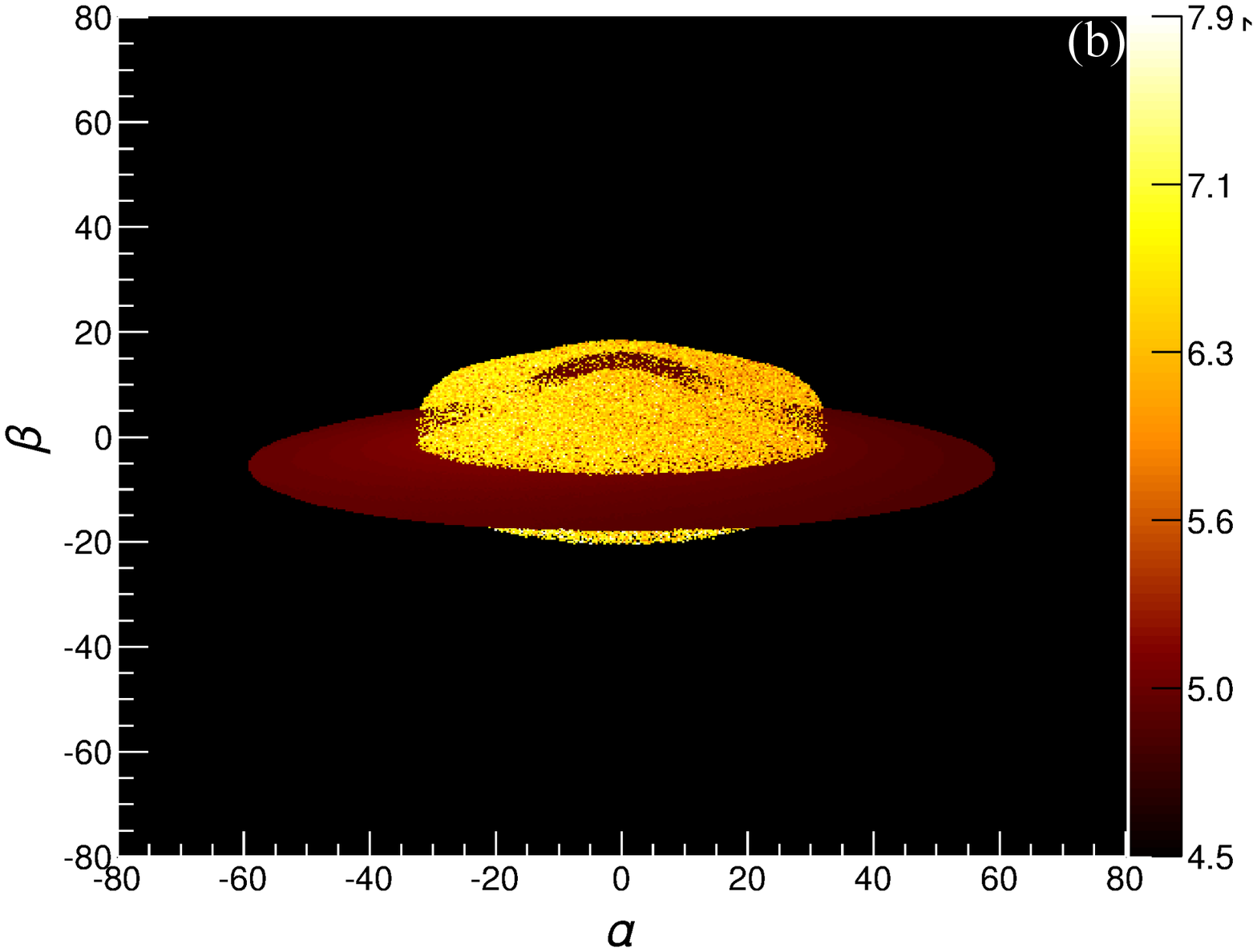}
    \includegraphics[width=0.31\textwidth,angle=0]{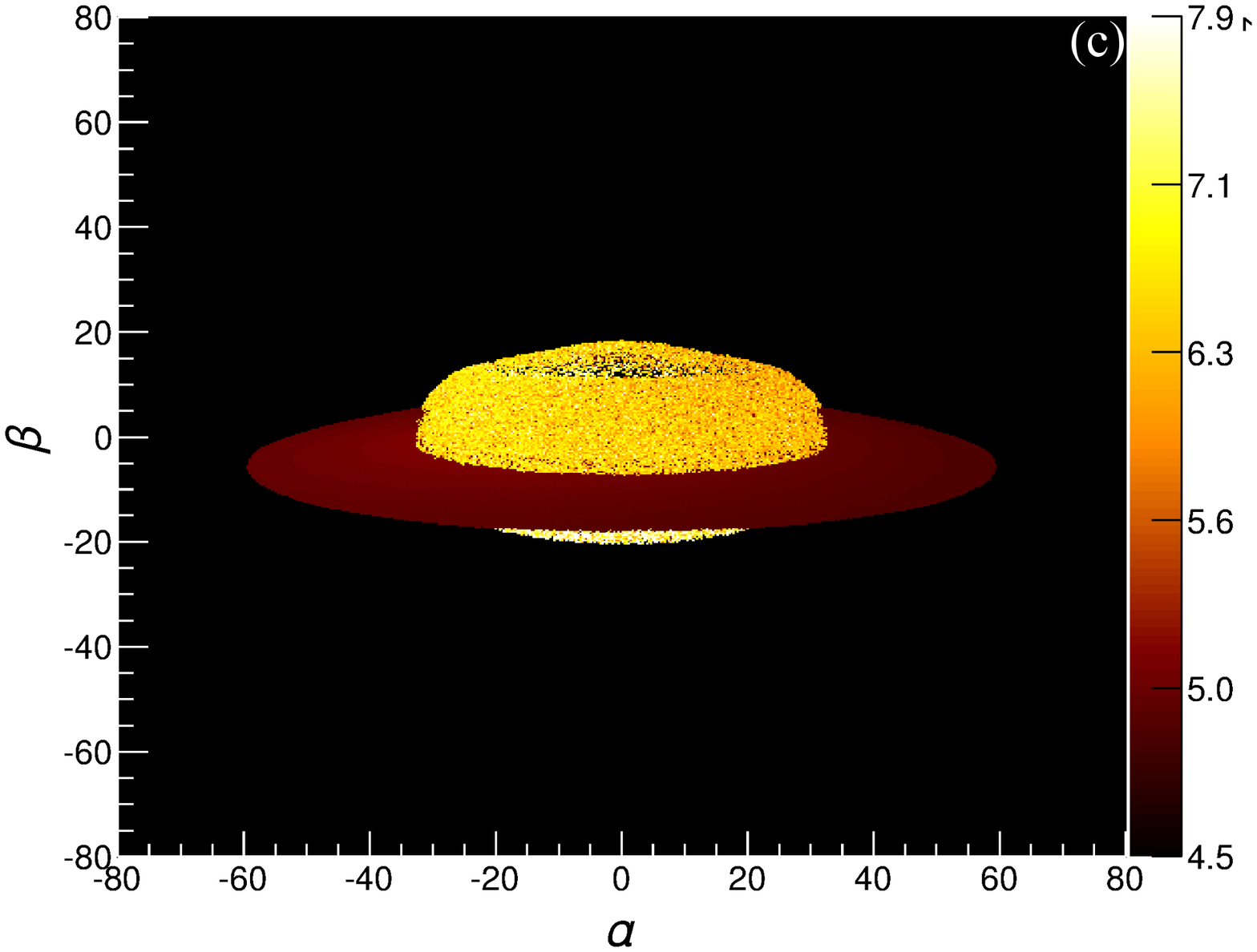}}\vskip 0.5cm
  \caption{Images seen by observers from $80^{\circ}$. Disk rate is kept constant at
$\dot{m_{d}}=0.00001$. The central temperature and density increases from left to right 
(Table 2): (a) $T_c=100$keV; $n_c=4\times 10^{16}$cm$^{-3}$; 
(b) $T_c=200$keV; $n_c=3\times 10^{17}$cm$^{-3}$; (a) $T_c=300$keV; $n_c=1.2\times 10^{18}$cm$^{-3}$. 
Color-bar is plotted in the $log(T_{obs})$ scale with $10^{4.5}-10^{7.9}~$ Kelvin is the
range fixed for each case.}
\end{figure}

\begin{table}
\centering
\caption{Variation of images with $T_c$ \& $n_c$}
\begin{tabular}{cccc}
  \toprule[1.5pt]
Figure & $\dot{m_{d}}$ & $T_c (keV)$ & $n_c (per~cm^3)$  \\
  \hline
10a          &  0.000001     & 100.0   & 4E16   \\
10b          &  0.000001     & 200.0   & 3E17   \\
10c          &  0.000001     & 300.0   & 1.2E18 \\ \hline
  \bottomrule[1.5pt]
\end{tabular}
\end{table}

The optical depth of the CENBOL medium quantifies the number of scatterings that 
should occur. In Fig. 10 (a-c), we have shown three cases where densities and temperatures of 
electron cloud were self consistently varied inorder to study CENBOL with three different optical depths. 
The parameters employed are given in Table 2. In the lower density
case, i.e., in Fig. 10a, photons from Keplerian disk have undergone very few
scatterings in the CENBOL region. The optical depth of the medium is so low that almost 
all the photons from the far side of the Keplerian disk are visible by the observer. Since the number of 
scattering is small, very less number of CENBOL photons are visible from lower side 
of the Keplerian disk also. Fig. 10b drawn for an intermediate optical depth, is same as the Fig. 9g.
Here some Keplerian photons cross over the electron cloud without scattering, especially in regions 
of lower optical depth. In Fig. 10c, we consider a relatively higher central density so that the 
optical depth is everywhere larger than unity. Here, almost all the photons which came from the other 
side of Keplerian disk participate in scattering and thus the Keplerian 
disk from far side becomes invisible.     

Photons from the disk and the CENBOL thus participate in the 
formation of the photon sphere. Although the contribution of such photons is very less 
in the spectrum, the measurement of size and shape of the photon sphere can determine 
both the mass and spin of the black hole candidate in future. We will discuss this elsewhere.

\section{Conclusions}

Photons emerging from the vicinity a black hole encounter the strongest possible gravitational fields.
So, spectral and temporal properties of a black hole should carry signatures of photon 
bending. Spectra of black hole candidates are primarily generated from a Keplerian 
disk (black-body) and a Compton cloud which is located in the inner region in the 
form of CENBOL (CT95). The physical parameters (accretion rates, shock location, shock 
strength, mass) of accretion process are extracted from the fitting of Two Component 
Advective Flow (TCAF) model (see Debnath et al. 2014; Mondal et al. 2014; Jana et al. 
2016; Molla et al. 2016; Chatterjee et al. 2016). So, imaging a black hole surroundings 
would necessarily mean imaging of the Keplerian disk along with the CENBOL. Ours is 
the first paper, which creates a composite image of both the components, namely, the Keplerian 
and the sub-Keplerian components which are accreting at different accretion rates after 
addition of realistic physical processes. We ignore the pre-shock region of 
sub-Keplerian component since it is optically very 
thin and concentrate only on the centrifugal barrier region which primarily does the inverse 
Comptonization to produce the power-law component. We ignored the radial velocity 
inside this barrier since the flow is highly subsonic till it reaches 
the inner sonic point located between the marginally bound and marginally stable 
orbits. We also present the corresponding 
spectra and images for various flow parameters. We find that multi-color black-body 
part of a spectrum becomes flat at higher inclination angles. The number of photons 
in higher energy bin ($10-100$~keV) increases with inclination angle. This is a clear 
effect of the focussing due to photon-bending. The spectra is harder at high inclination angles. 

Our work has, for the first time, produced the images and also realistic spectra from 
a two component advective flow around a non-rotating black hole. Due to intrinsic 
nature of our Monte-Carlo simulation, we can also not only study the effects of 
bending on the timing properties, but also determine the time and phase lag properties. 
Similarly effects of spin would be very important and we would like to identify those features
which are especially on spin dependent. This is outside the scope of this paper 
and would be reported elsewhere.

\section{Acknowledgements}

The work of AC is supported by Ministry of Earth Science (MoES), India.

%%%%%%%%%%%%%%%%%%%%%%%%%%%%%%%%%%%%%%%%%%%%%%%%%%%%%%%%%%%%%%%%%%%%%
{}
%%%%%%%%%%%%%%%%%%%%%%%%%%%%%%%%%%%%%%%%%%%%%%%%%%%%%%%%%%%%%%%%%%%%%
\end{document}